
\magnification 1200
\baselineskip= 0.7 truecm
\nopagenumbers
\footline={\hss\tenrm\folio\hss}
\hsize=16.0 truecm
\vsize=22. truecm
\voffset=1.5 truecm
\tolerance=3000
\centerline {\bf
A PROCEDURE TO CALIBRATE A MULTI-MODULAR TELESCOPE }
\vskip .45 truecm
\centerline {\rm P.F.MASTINU, P.M.MILAZZO, M.BRUNO, M.D'AGOSTINO}
\centerline {\it Istituto Nazionale di Fisica Nucleare, Sezione di Bologna e}
\centerline {\it Dipartimento di Fisica dell'Universit\`a di Bologna}
\centerline {\it Via Irnerio 46, 40126 Bologna, Italy}
\vskip 1.0 truecm
\centerline{\bf Abstract}
\vskip .5 truecm
A procedure has been developed for the charge, mass and energy calibration
of ions produced in nuclear heavy ion reactions. The charge and mass
identification are based on a $\Delta$E-E technique. A computer code determines
the conversion from ADC channels into energy values, atomic number and mass
of the detected fragments by comparing with energy loss calculations
through a minimization routine. The procedure does not need
prior measurements with beams of known energy and charge.
An application of this technique to the calibration of the MULTICS apparatus
is described.
\vfill\eject
\centerline{\bf 1. Introduction }
\vskip .5 truecm
In the last several years new detectors with large solid angles and geometrical
efficiencies have been developed
to investigate heavy ions reactions at intermediate energies
(10 MeV/u-1 GeV/u) [1-4].
The new experimental apparata allow the simultaneous detection of
energy, emission angle, atomic number and mass of several
fragments by using a large number of multimodular telescopes.
To extract this kind of information from the detector signals
long and tedious calibrations are usually required. This is due to the high
number of different detectors (ionization chambers, semiconductors and
scintillators) and to the large number of nuclear species in a wide energy
range which are produced in the reactions.
It is important to note that these large experimental
apparata allow relatively fast data collection and are able to detect also
rare events due to their high detection efficiency. On this basis
it is possible to measure with different projectile+target systems
at different beam energies with the same experimental setup in the available
beam-time.
Furthermore since different
amplification gains of the electronic chain may be needed during a single
experiment (because of different beam
energy and projectile+target combination) a longer calibration procedure may
result.\par
Part of the beam-time is typically used for the collection of known-energy
experimental points (as Rutherford diffusion on targets or direct exposition
to low-intensity beam) from which
the detectors calibration is then determined.\par
In this paper we will present
the implementation of a new time-saving calibration method
to minimize the time dedicated to detector calibration, thus allowing the
collection of more data, hence smaller statistical uncertainties.
Furthermore in all the cases when known-energy beams are not available or
it is not possible, using elastic scattering, to cover all
detectors and telescopes, the procedure we will describe
allows one to obtain the angular coefficient (slope) and constant term
(offset) of the straight-line calibration. \par
It must be considered also that possible instabilities of the electronics
may cause amplification drift making it difficult to use the same set
of calibration data in two measurements made at different times.
It is then extremely useful to have a calibration procedure which can be
applied to each run independently.
In order to optimize the experimental potential of a quick data
analysis a fast data reduction must also be implemented along
with fast data collection.\par
Last but not least the procedures commonly used for the charge (mass)
calibration are based on graphical cuts or on Particle Identification Functions
(P.I.F.) which, because of their manual nature,
automatically introduce a limitation into the calibration.
On the contrary a minimization procedure makes it possible
to reduce the human "subjectivity" of the results and to achieve
higher precision.\par
With this technique pre-calibration procedures can be skipped and human
participation is required only to check the results. This technique is made
possible by the high computational abilities of modern computers.\par
\vskip 1.0 truecm
\centerline {\bf 2. The calibration procedure}\par
\vskip 0.5 truecm
\rm
The calibration procedure here described may be applied to the
multi-modular detectors
with resolution sufficient to separate the curves relative to single atomic
number (mass) in a ($\Delta E, E_{res}$) matrix, where $\Delta E$ and $E_{res}$
are, respectively, the energy losses in two successive detectors of an
incident ion with total energy $E=\Delta E+E_{res}$.\par
Every single section in the program compares the data to the energy loss
calculations.
Specific energy loss (${{dE}\over {dx}}$) of a charged particle in matter
depends
on the characteristics of the incident ion (mass, charge and energy) and of
the absorber medium (volumetric density and atomic number) and is well
described by the classical Bethe-Bloch formula [5].
Energy loss calculation, based on the Anderson studies [6-7],
are able to reproduce experimental
data with good accuracy in a large energy and atomic species spectra. This
can be seen in fig. 1 where the energy loss curves overlap
experimental data relative to ion species of beams with known energy.\par
The calibration program consists of three different sections:\par
\item{  i)} a first introductory section that deals with energy losses in
detectors;
\item{ ii)} a second section dedicated to energy calibration; in this section
the offsets and amplification gains of electronic chain are computed
for the various detectors of each telescope;
\item{iii)} a final event-by-event calibration in energy, charge and mass. \par
As an example, we will describe the application of the procedure to a
telescope of the MULTICS array, which consists of a ionization chamber (IC), a
500 $\mu m$ thick
silicon (SI), a 4 $mm$ lithium drifted silicon (SILI) and a $CsI(Tl)$
scintillator (CSI).\par
{\bf 2.1 Energy loss tables }\par
Part $i)$ deals with the preparation of tables containing the data for the
energy loss of various ions at several energies in each pair of
$\Delta E-E_{res}$ detectors. For an accurate calibration it is extremely
important
to know with good precision each $\Delta E$ detector active thickness and dead
layer. If the telescope is not able to determine the ion mass $A$, the most
probable isotope associated with the
atomic number $Z$,
$A=a\cdot Z+b\cdot Z^2$ ($a=2.08$, $b=0.0029$
for fragments produced in heavy ion reactions [8-9], $b=0.0059$ for the most
probable isotopes on the periodic table) can be used in the energy loss
calculation.
The differences in the $b$ values take into account the fact that the reaction
products
tend to be lighter than the most probable isotopes in nature because of the
preferential emission of neutrons.\par
The aim of this first procedure
is to obtain an analytical expression for the energy-loss tabulated points.
This can be done also before the experiment
and remains valid during all the measurements done using the same
detectors.\par For every given ion the following
analytic form connects the various energy losses in the
telescope:\par
$$ \Delta(E)_{Z,A}=f_{Z,A}(E_{res})=-d_1\cdot E_{res}^{-d_2}\cdot exp(1-{E_{res
} ^{d_3}\over d_4})\eqno(1)$$
where $d_1$, $d_2$, $d_3$, and $d_4$ are the parameters to be
determined for each $(Z$, $A)$ pair in the energy range
$0 \le E_{res} \le E_{MAX}$.\par
In figure 2a we show the good agreement between the curves
$\Delta (E)_{Z,A} =f_{Z,A}(E_{res})$ and the
points calculated through eq.(1) for different values of atomic number $Z$.\par
{\bf 2.2 Energy calibration of the telescopes }\par
After the determination of the $f_{Z,A}(E_{res})$ parameters, the program
then deals with the energy calibration
of each telescope.\par
A set of points is extracted for every $\Delta E-E_{res}$ matrix.
Data sampled on the curves are shown in fig. 2b; one has to assign arbitrarily
a temporary $Z$
value ($Z_{TEMP}$) to each curve in the right order given by subsequent curves.
With the program the correct
correspondence between $Z$ and the relative curve will be done
through the option that allows to add or subtract a constant value $Z_{plus}$
to each $Z_{TEMP}$. \par
The marked points shown in fig. 2b are extracted from
the $\Delta E-E_{res}$ matrix and the coordinates are put in a table.
Each of these points is characterized by its $\Delta E(Ch)$ and
$E_{res}(Ch)$ coordinates (expressed in channels)
and by a $Z_{TEMP}$ value. Once the first is fixed,
a $Z_{TEMP}+1$ value must be assigned to the next curve
and so on. The angular coefficients and the
known-terms of the
calibration curves for the various detectors in the telescope are treated
as parameters in the minimization routine (we use the MINUIT D505 routine
from the CERN Program Library).\par
Through a minimization process the program determines the angular coefficients
and the offsets of the energy-channel curves
that yield the best agreement between sampled points and the
energy-loss curves from eq. (1). \par
The method consists of minimizing the
distance between the sampled points and the $\Delta (E)_{Z,A}
=f_{Z,A}(E_{res})$
curves associated with the relative $Z$ value ($Z_{TEMP}$).\par
The $\chi^2$ value for each event is given by the squared difference
between the value of the temporary calibration of an
experimental $\Delta E$ value and the value predicted from the energy-loss
curves corresponding to a chosen $Z_{TEMP}$ value and to an
$E_{res}$-experimental-calibrated-value.
The code makes a comparison between the different
$\Delta E$ values, that correspond to a fixed value $E_{res}$ for the
temporary assigned $Z_{TEMP}$ value.\par
The final parameters, obtained by the $\chi^2$ fit, give the offsets and
angular coefficients best able to match the
energy-loss $\Delta (E)_{Z,A} =f_{Z,A}(E_{res})$ curves to sampled data
(therefore to experimental data) for a particular choice of the
assigned $Z_{TEMP}$-values.\par
Using $Z_{plus}$ as a variable
the minimization procedure is repeated with different $Z$ values
assigned to the sampled data;
the $\chi^2$ shows an evident minimum
corresponding to the correct assignment of the $Z$ value to each
curve. \par
In fig. 2c we show a $\Delta E-E_{res}$ matrix calibrated with the
coefficients obtained from the minimization method discussed above.
The $\Delta (E)_{Z,A} =f_{Z,A}(E_{res})$ curves for different $Z$
values overlap the experimental data.\par
It should be noted that when dealing with a multimodular telescope the
calibration of the middle
detectors (those that act as passing and stopping detectors)
applies heavy constraints to the calibration quality if one wants good
$f_{Z,A}(E_{res})$ fits on two different matrices, since the
choice of offset and slope values for the middle
detector impose a constraint on both the matrices.\par
The good capability of the method is clearly shown
in fig. 3, where the curves corresponding to different $Z$ values
are overlapped onto the experimental calibrated data. In this figure the
abscissa $E_{res}$ is obtained from the sum of the output of the two
following detectors.\par
{\bf 2.3 Event by event calibration }\par
After the energy calibration the atomic number $Z$ has to be assigned to
each experimental point, event by event. \par
Starting from the energies, obtained as described before or from
energy-known beams, the program estimates the distance of the experimental
point to the
curve $\Delta$E=$f_{Z,A}(E_{res})$ for all the possible $Z$ values.
The shortest distance between the ($E_{res}$, $\Delta$E) point and
a curve $f_{Z,A}(E_{res})$ determines the choice of the appropriate
assignment of the $Z$-value.
The $Z$-value is evaluated in the matrix corresponding to the
detector which works as stop detector for that particular event. This
detector can be determined as the one where
the subsequent detector measures zero energy. \par
We have applied the same calibration procedure to a $CsI(Tl)$ scintillator
through a calibration curve which is able to reproduce
the light-response of the scintillator as a function of the energy and charge
of the incident ion [10-11]. By implementing this function in
a routine we could consider the scintillator in a manner similar to the
other linear detectors. \par
The whole calibration procedure has been tested by comparing it with results
obtained through standard methods (e.g. P.I.F. for
the charge calibration). The precision of the calibration
has also been checked with the calibration obtained from known
energy experimental points which have been measured by low-intensity beam
impinging directly on the detectors. The excellent consistency of the results
shows the quality of our procedure. \par
Exploiting the possibility of accelerating simultaneously ions with a
charge/mass
ratio roughly constant, several "cocktail" beams have been obtained
at the Superconducting Cyclotron of the Michigan State University
(18 different ion species have been accelerated). With this "cocktail" beams a
very accurate calibration was obtained independently.
We checked the data calibrated with the previously described
procedure with data calibrated by the cocktail beams.
A very good agreement between the two calibration methods was found.
The excellent
agreement found for the calibration parameters obtained with these two
different methods gives added credibility to
the capability of the technique described here.\par
In fig. 2c the agreement between the experimental calibrated data
with the procedure discussed above and the corresponding energy-loss
curves confirms the accuracy of the method. \par
\vskip 1.0 truecm
\centerline{\bf Conclusions} \par
\vskip 0.5 truecm
\rm
The advantages obtained with this automatic calibration procedure may be
summarized in the following points.\par
\item{-} Known-energy beams or target elastic diffusion calibration
are not needed. This results in more beam time available for experimental
data collection.
\item{-} The time dedicated to offline calibration is greatly reduced.
\item{-} Results are very accurate, comparable with those obtained
with procedures based on the collection of a large number of known-energy
experimental points and using P.I.F.
\item{-} The procedure can be easily extended to any experimental apparata with
multi-modular
telescopes based on $\Delta E-E_{res}$ technique
for the evaluation of the ion atomic number and/or mass.\par
\vskip 1.0 truecm
\centerline{\bf Acknowledgements}\par
\vskip 0.5 truecm
\rm
The authors wish to thank N. Colonna, N. Moggi and M. Pavan for the useful
comments and help.\par
This work was supported in part by the Italian Ministry of University and
Scientific Research.
\vfill\eject
\centerline{\bf References}\par
\item {[1] } I.Iori et al. Nucl. Instr. and Meth. A325 (1993) 458
\item {[2] } R.T. de Souza et al. Nucl. Instr. and Meth. A295 (1990) 109
\item {[3] } J. Pouthas et al. Nucl. Instr. and Meth. A357 (1995) 418
\item {[4] } U.Lynen et al. Gesellschaft f\"ur Schwerionenforschung Report n.
GSI-02-89
\item {[5] } H.A.Bethe Ann. Physik 5,325 (1930)
\item {[6] } H.H.Anderson and J.F.Ziegler Stopping power and ranges in
all elements, Pergamon Press (1977)
\item {[7] } F.Hubert, R.Bimbot and H.Gauvin, Nucl. Instr. and Meth. B36 (1989)
357
\item {[8] } A.Gavron, Phys. Rev. C21 230 (1986)
\item {[9] } R.J.Charity et al. Phys. Rev. Lett. 56 (1986) 1354
\item {[10] } N.Colonna et al. Nucl. Instr. and Meth. A321 (1992) 529
\item {[11] } P.F.Mastinu, P.M.Milazzo, M.Bruno, M.D'Agostino, L.Manduci.
Nucl. Instr. and Meth. A341 (1994) 663
\vfill\eject
\bf{Figure captions}\par
\rm
\bf{Fig. 1} \rm Matrix $\Delta E$ vs $E_{res}$ from direct exposure to
energy-known beams (500$\mu
m$ Silicon as $\Delta E$ detector (SI) and $4 mm$ of Lithium drifted Silicon
as stop detector (SILI)). \par
\bf{Fig. 2} \rm a) Curves $\Delta E-E_{res}$ for the matrix $\Delta E$
(Silicon,
500$\mu m$ thick, SI) vs $E_{res}$ ($4 mm$ Lithium drifted Silicon, SILI). The
points come from the output of the ENLOSS code while the overlapped curves come
from the fit with the analytic function (1). One can see the good capability of
the function to fit the large range of data.
b) Example of sampled point. It is not
important to sample all the curves. The number of sampled points is about 90.
c) Calibrated data overlapped by energy-loss curves from $Z=2$ up to
$Z=22$. One can see the good agreement
between the curves and the experimental calibrated points, demonstrating the
good accuracy of the calibration method. \par
\bf{Fig. 3} \rm Curves $\Delta E-E_{res}$ for the matrix $\Delta E$
(Ionization chamber, 8.5 $cm$ long, $CF_4$ gas filled at a pressure of 90
$mbar$, IC) vs $E_{res}$ (Silicon + Lithium drifted Silicon, SI+SILI)
for $Z=8$ up to $Z=22$ in step of 2; energy loss curves are overlapped.
In this case Silicon act as passing and stopping detectors.
\vfill\eject\bye